\documentclass[aps,preprint,amsmath,amssymb]{revtex4-1}
\usepackage{slashed}
\usepackage{graphicx}
\usepackage{color}
\newcommand{\nn}{\nonumber}
\newcommand{\bd}{\begin{document}}
\newcommand{\ed}{\end{document}}
\newcommand{\bc}{\begin{center}}
\newcommand{\ec}{\end{center}}
\newcommand{\be}{\begin{eqnarray}}
\newcommand{\ee}{\end{eqnarray}}
\newcommand{\ba}{\begin{array}}
\newcommand{\ea}{\ed{array}}
\newcommand{\strich}[1]{#1  \! \! \slash}
\newcommand{\eqn}{\global\def\theequation}
\newcommand{\sw}{sin^2 \theta_W}
\newcommand{\fbd}{f_B}
\renewcommand{\thefootnote}{\alph{footnote}}
\newcommand{\se}{\section}
\newcommand{\sse}{\subsection}
\newcommand{\bi}{\bibitem}
\def\figcap{\section*{Figure Captions\markboth
     {FIGURECAPTIONS}{FIGURECAPTIONS}}\list
     {Figure \arabic{enumi}:\hfill}{\settowidth\labelwidth{Figure 999:}
     \leftmargin\labelwidth
     \advance\leftmargin\labelsep\usecounter{enumi}}}
\let\endfigcap\endlist \relax
\def\reflist{\section*{References\markboth
     {REFLIST}{REFLIST}}\list
     {[\arabic{enumi}]\hfill}{\settowidth\labelwidth{[999]}
     \leftmargin\labelwidth
     \advance\leftmargin\labelsep\usecounter{enumi}}}
\let\endreflist\endlist \relax

\def\Journal#1#2#3#4{{#1} {{\bf #2},} {#4} {(#3)}}
\def\NCA{Nuovo Cimento}
\def\NIM{Nucl. Instrum. Methods}
\def\NIMA{{Nucl. Instrum. Methods} A}
\def\NP{{Nucl. Phys.} }
\def\NPB{{Nucl. Phys.} B }
\def\NPA{{Nucl. Phys. A}}
\def\PLB{{Phys. Lett.}  B}
\def\PL{{Phys. Lett.}}
\def\PPSA{{Proc. Phys. Soc.} A}
\def\PRP{{ Phys. Rep.}}
\def\PRL{ Phys. Rev. Lett.}
\def\PR{{Phys. Rev.}}
\def\PRD{{Phys. Rev.} D}
\def\PRC{{Phys. Rev.} C}
\def\ZP{{Z. Phys.}}
\def\ZPC{{Z. Phys. C}}
\def\EPJ{{Eur. Phys. J.}}
\def\EPJC{{Eur. Phys. J.} C}
\def\ZPA{{Z. Phys.} A}
\def\MPL{{Mod. Phys. Lett.}}
\def\MPLA{{Mod. Phys. Lett.} A}
\def\CPC{Comput. Phys. Commun.}
\def\JHEP{{J. High Energy Phys.}}
\def\JPG{{J. Phys. G.}}
\def\SJNP{Sov. J. Nucl. Phys.}
\def\NCA{ Nuovo Cimento}
\def\NIM{ Nucl. Instrum. Methods}
\def\NIMA{{ Nucl. Instrum. Methods} A}
\def\NP{{ Nucl. Phys.}}
\def\ANP{{Adv. Nucl. Phys.}}
\def\CPC{{Comput. Phys. Commun.}}

\begin{document}
\title{Semileptonic Decays of $\Lambda \to p \ell^{-} \bar{\nu}_{\ell}$ in the Light-Front Dynamics}

\author{Chong-Chung Lih\footnote{cclih123@gmail.com} and 
Chao-Qiang Geng\footnote{cqgeng@ucas.ac.cn}}
\affiliation{
Chongqing University of Posts and Telecommunications, 
Chongqing, 400065, China\\
School of Fundamental Physics and Mathematical Sciences,
Hangzhou Institute for Advanced Study, UCAS, Hangzhou 310024, China}

\date{\today}

\begin{abstract}

We investigate the exclusive semileptonic decays of 
$\Lambda \to p \ell^{-} \bar{\nu}_{\ell}~(\ell=e,\mu)$  within the standard model 
using the light-front quark model. 
The transition form factor behaviors of $\Lambda \to p$ 
are obtained from the effective treatment of
nonvalence contributions in addition to the valence ones
in the Drell-Yan-West frame due to the Bethe-Salpeter formalism.
Based on these form factors, we obtain that the branching ratios of 
$\Lambda\to p e^{-} \bar{\nu}_{e}$ and $p \mu^{-} \bar{\nu}_{\mu}$, 
including nonvalence contributions, 
are around $8.32\times 10^{-4}$ and $1.31\times 10^{-4}$,  
which are consistent with the latest measurements from the BESIII Collaboration, respectively. 
Our results indicate that nonvalence contributions can play a non-negligible role 
in the semileptonic baryon decays within the light-front framework.

\end{abstract}

\maketitle

\se{Introduction}

Semileptonic decays of mesons and baryons 
have long provided valuable insights for testing the Standard Model (SM) 
and play a crucial role in understanding the interplay between 
strong and weak interactions. In particular, they can be used 
to conduct rigorous tests of the Cabibbo-Kobayashi-Maskawa (CKM) 
mixing elements within the SM~\cite{cp1,cp2,cp3}. 
The ratio of the semileptonic decay rates, 
$R_{\mu e}^{B_1B_2}=\Gamma(B_{1}\to B_{2} \mu^{-} \bar{\nu}_{\mu})
/\Gamma(B_{1}\to B_{2}  e^{-} \bar{\nu}_{e})$, 
between the muon and electron decay channels has been extensively 
studied in the literature~\cite{QCDSR1,QCDSR2,QCDSR3,Cabibbo,quarkmodel1,quarkmodel2,1/Nc,Lattic,Nonst,LQCD}. 
Recently, the BESIII~\cite{BESIIIV,BESIIIX} and LHCb~\cite{LHCb1} Collaborations have
reported updated results on the $\Lambda\to p\ell^{-} \bar{\nu_{\ell}}$ decays,  given by
\be
{\cal B}(\Lambda\to p\mu^{-} \bar{\nu_{\mu}})&=&(1.48\pm0.21\pm0.08) 
\times 10^{-4}\,,  \nonumber\\
{\cal B}(\Lambda\to p e^{-} \bar{\nu}_{e})&=&(8.16\pm0.22\pm0.14) 
\times 10^{-4}\,,  \nonumber\\
R_{\mu e}^{\Lambda p}|_{{\rm exp}}&=&0.181\pm0.028\,,
\ee
and
\be
{\cal B}(\Lambda\to p\mu^{-} \bar{\nu_{\mu}})&=&(1.462\pm0.1) 
\times 10^{-4}  \nonumber\\
R_{\mu e}^{\Lambda p}|_{{\rm exp}}&=&0.175\pm0.012\,,
\ee
respectively.
As pointed out in Ref.~\cite{Nonst},
the ratio of $R_{\mu e}^{\Lambda p}$
is not only independent of $V_{us}$ and the lifetime of $\Lambda$, but is free of hadronic form- factor uncertainties at the next-to-leading order in the SM.
Explicitly, the authors~\cite{Nonst} predicted that
$R_{\mu e}^{\Lambda p}=0.153\pm0.008$, which 
was recently updated to be $0.162\pm0.004$ in Ref.~\cite{LQCD} based on the data in the PDG~\cite{PDG}.
 From a theoretical perspective, 
this type of decay process involves an $s\to u$ transition mediated by a
virtual $W$ boson, making it a useful process to study weak 
interactions and hadron structures.

The transition form factor is a fundamental property of hyperons that 
describes the dynamic behavior of the transition between two states. 
The hadron part of the weak decay amplitude of a hyperon into a proton is 
described by the transition matrix element 
$\langle p|\bar{u}\gamma^{\mu} (1-\gamma_{5}) s|\Lambda \rangle$, 
which is parameterized by six form factors of 
$f_{i}(q^{2})$ and $g_{i}(q^{2})$ ($i = 1, 2, 3$), given by
\be
&&\langle {\cal B}_{p}(P^{\prime},S^{\prime}=
\frac{1}{2},S_{z}^{\prime})|\bar{u}\gamma^{\mu}
(1-\gamma_{5})s|{\cal B}_{\Lambda}(P,S=\frac{1}{2},S_{z})\rangle \nonumber\\
&& =  \bar{u}_{\alpha}(P^{\prime},S_{z}^{\prime})
\Big[\gamma^{\mu} f_{1}(q^{2})
+i \frac{f_{2}(q^{2})}{M_{\Lambda}}\sigma^{\mu\nu}q_{\nu}
+\frac{f_{3}(q^{2})}{M_{\Lambda}} q^{\mu}\Big]
u(P,S_{z})\nonumber\\
&&\quad-\bar{u}_{\alpha}(P^{\prime},S_{z}^{\prime})
\Big[\gamma^{\mu} g_{1}(q^{2})
+i \frac{g_{2}(q^{2})}{M_{\Lambda}}\sigma^{\mu\nu}q_{\nu}
+\frac{g_{3}(q^{2})}{M_{\Lambda}} q^{\mu}\Big]\gamma_{5}u(P,S_{z}),
\label{transitionVA}
\ee
where $q=P-P^{\prime}$, $P^{(\prime)}$ and $S_{z}^{(\prime)}$ 
represent the momentum and spin of $\Lambda$ (proton), respectively. 
It follows that these form factors encode the nonperturbative 
strong-interaction dynamics of the baryonic transition.

In this work, we employ the light-front quark model (LFQM) based on 
the light-front (LF) quantization~\cite{lf1,lf2,BS1,BS2,BS3,lf3} 
to calculate the hadronic form factors in Eq.~(\ref{transitionVA}) 
and evaluate the decay rates of 
$\Lambda\to p\,\ell^{-} \bar{\nu}_{\ell}~(\ell = e,\mu)$ within the SM. 
From a phenomenological perspective, the LF formalism provides a simple, 
nonperturbative, and relativistic framework for calculating hadronic 
form factors. 
In this paper, the LFQM incorporates a diquark 
model~\cite{diquark,diquark2}, in which a baryon consists 
of an active quark and a spectator diquark, thereby allowing the non-perturbative 
interaction between the two light quarks to be effectively 
incorporated into the diquark mass. 
This approximation effectively reduces the baryonic system to a two-body problem, 
significantly reducing the computational burden. 
The LFQM~\cite{lf3} framework adopted in this work has several noteworthy features 
compared to other LFQM~\cite{lf4,LFQM1} analyses: 
(1) We evaluate the six form factors in the spin $1/2 \to 1/2$ 
transition matrix element in LF coordinates.
(2) We analyze the form factors in a timelike space and extend 
them to the entire physical region, thereby obtaining the form factors 
of the  baryon semileptonic decays. 
(3) We use the recent methods proposed in Refs.~\cite{BS1,BS2,BS3,lf3} with an 
effective treatment based on the Bethe-Salpeter (B-S) formalism (see~\cite{BS1,BS2,BS3,LFg}) 
for addressing the higher-order Fock contributions to the 
form factors in the $q^{+}>0$ coordinate system.

In the LF framework, the $``\mu"$ index 
in the V-A current takes the $``+"$ component, 
while the  calculation is performed in the range with $q^{+}\neq0$, 
i.e. in the timelike domain ($q^{2} > 0$).
From a physical perspective, since semileptonic decays probe the 
in physical timelike region, it is preferable to evaluate the form factors 
in the $q^2 > 0$ domain without relying solely on analytic continuation. 
Similarly, when conducting studies in the timelike domain, 
contributions from nonvalence Fock states may also appear due to
the inclusion of higher-order Fock states in the wave function, 
in addition to valence states~\cite{BS1}. In other words, 
we will inevitably encounter nonvalence diagrams arising from 
the creation of quark-antiquark pairs. Note that, 
effective methods for handling nonvalence contributions 
in meson-meson scattering processes have already been established 
in the literature~\cite{BS1,BS2,BS3,LFg}. 

This paper is organized as follows.
In Sec. II, we present the framework for the baryonic semileptonic decays 
of $\Lambda\to p\,\ell^{-} \bar{\nu}_{\ell}$.
Our numerical results and discussions are given in  Sec. III. 
We conclude in Sec. IV.

\section{framework for the semileptonic decays of $\Lambda\to p\,\ell^{-} \bar{\nu}_{\ell}$}

In this section, we present the theoretical framework for the semileptonic decay 
of $\Lambda\to p\,\ell^{-} \bar{\nu}_{\ell}$ within the LFQM. 
The decay amplitudes are governed by the effective weak Hamiltonian describing 
the  transition of $s\to u$ at quark-level,
\be
H_{eff}(\Lambda\to p\,\ell^{-} \bar{\nu}_{\ell})= {G_F\over \sqrt{2}}V_{us}
\langle {\cal B}_{p}(P^{\prime},S^{\prime}=
\frac{1}{2})|\bar{u}
\gamma^{\mu}(1-\gamma_5)s|{\cal B}_{\Lambda}(P,S=\frac{1}{2})\rangle
\,\bar{\nu}_{\ell}\gamma^{\mu}(1-\gamma_5)\ell\,.~
\label{he2}
\ee
We note that the discussion in this section can be viewed as a generic study of ${\cal B}^{\prime}(P^{\prime},S^{\prime})\to {\cal B}(P,S=\frac{1}{2})\ell^{-} \bar{\nu}_{\ell}$
for any ${\cal B}{\prime}(1/2) \to {\cal B}(1/2)$ transition.

\sse{Helicity amplitudes}

The transition matrix elements in Eq.~(\ref{transitionVA}) can be applied to
the helicity amplitudes,
given by~\cite{lf11,HA1,HA2}
\begin{eqnarray}
H^{V(A)}_{\lambda_{p} \lambda_{W}}
\equiv\langle p|(\bar u s)_{V(A)}|\Lambda\rangle
\varepsilon^\mu_W\,,
\label{helicityA}
\end{eqnarray}
where $\varepsilon^\mu_W$ is the polarization of the $W$ boson,
and $\lambda_{p}=\pm 1/2$
represent the helicity states of the proton.
Based on the helicity conservation,
$\lambda_{\Lambda}=\lambda_{p}-\lambda_{W}$ is held.

The helicity amplitudes are related to the form factors through the
following expressions:
\begin{align}
H_{\frac{1}{2},t}^{V} & =-i\frac{\sqrt{Q_{+}}}{\sqrt{q^{2}}}
\left((M_{\Lambda}-M_{p})f_{1}+\frac{q^{2}}{M_{\Lambda}}f_{3}\right),\nonumber \\
H_{\frac{1}{2},0}^{V} & =-i\frac{\sqrt{Q_{-}}}{\sqrt{q^{2}}}
\left((M_{\Lambda}+M_{p})f_{1}-\frac{q^{2}}{M_{\Lambda}}f_{2}\right),\nonumber \\
H_{\frac{1}{2},1}^{V} & =-i\sqrt{2Q_{-}}\left(f_{1}
-\frac{M_{\Lambda}+M_{p}}{M_{\Lambda}}f_{2}\right),\nonumber \\
H_{\frac{1}{2},t}^{A} & =-i\frac{\sqrt{Q_{-}}}{\sqrt{q^{2}}}
\left((M_{\Lambda}+M_{p})g_{1}-\frac{q^{2}}{M_{\Lambda}}g_{3}\right),\nonumber \\
H_{\frac{1}{2},0}^{A} & =-i\frac{\sqrt{Q_{+}}}{\sqrt{q^{2}}}
\left((M_{\Lambda}-M_{p})g_{1}+\frac{q^{2}}{M_{\Lambda}}g_{2}\right),\nonumber \\
H_{\frac{1}{2},1}^{A} & =-i\sqrt{2Q_{+}}
\left(g_{1}+\frac{M_{\Lambda}-M_{p}}{M_{\Lambda}}g_{2}\right).
\end{align}
where the subscript ``$t$" denotes $H_{\frac{1}{2},t}^{V(A)}$
from the temporal component of the current of $(\bar u s)_{V(A)}$, 
$q^{2}$ is the lepton pair invariant mass,
$Q_{\pm}=(M_{\Lambda} \pm M_{p})^{2}-q^{2}$, 
and $M_{\Lambda}$ ($M_{p}$) is the mass of the parent (daughter) baryon.
The negative helicity amplitudes are defined by
\begin{equation}
H_{-\lambda_{p},-\lambda_{W}}^{V}=H_{\lambda_{p},
\lambda_{W}}^{V}\quad\text{and}\quad H_{-\lambda_{p},-\lambda_{W}}^{A}
=-H_{\lambda_{p},\lambda_{W}}^{A},
\end{equation}
while the helicity ones for the left-handed current are obtained as
\begin{equation}
H_{\lambda_{p},\lambda_{W}}=H_{\lambda_{p},
\lambda_{W}}^{V}-H_{\lambda_{p},\lambda_{W}}^{A}.
\end{equation}
Consequently,
the differential decay widths of the semileptonic processes read
\begin{align}
\frac{d\Gamma_{L}}{dq^{2}}&=&\frac{G_{F}^{2}|V_{us}|^{2}}{(2\pi)^{3}}
\frac{(q^{2}-m_{\ell}^{2})^{2}\,\rm p_{cm}}{24M_{\Lambda}^{2}\,q^{2}}
\bigg\{\left(1+\frac{m_{\ell}^{2}}{2 q^{2}}\right)\left[|H_{\frac{1}{2},0}|^{2}+|H_{-\frac{1}{2},0}|^{2}
+|H_{\frac{1}{2},1}|^{2}+|H_{-\frac{1}{2},-1}|^{2}\right]\nonumber \\
&&+\frac{3 m_{\ell}^{2}}{2 q^{2}}\left(|H_{\frac{1}{2},t}|^{2}
+|H_{-\frac{1}{2},t}|^{2}\right)\bigg\}\,,
\label{Dwidth}
\end{align}
where $\rm p_{cm}$$ =\sqrt{Q_{+}Q_{-}}/2M_{\Lambda}$. 

\sse{Form factors in light-front quark model}

In order to calculate the form factors in the LFQM,
we treat the baryon as a bound state described in the quark-diquark picture of $q_1$ and $q_{2,3}$, 
where $q_{2,3}$ are combined into a single diquark, expressed as $q_{[2,3]}$.
Explicitly, the baryon bound state with the total momentum
$P$ and spin $S$ can be written as~\cite{diquark2,lf3,lf4}
\be
|{\cal B}(P,S,S_{z})\rangle & = & \int\{d^{3}p_{1}\}
\{d^{3}p_{2}\}2(2\pi)^{3}\delta^{3}(P-p_{1}-p_{2})\nonumber \\
&  & \times\sum_{\lambda_{1},\lambda_{2}}\Psi^{SS_{z}}
(p_{1},p_{2},\lambda_{1},\lambda_{2})|q_{1}(p_{1},\lambda_{1})[q_2, q_3]
(p_{2},\lambda_{2})\rangle\,,
\label{boundstate}
\ee
where $q_{1}=s$ or $u$ denotes the active quark corresponding to
$\Lambda$ or $p$, $[q_2, q_3]$ represents the diquark,
$\Psi^{SS_{z}}$ corresponds to the momentum-space wave function and
$p_{1,2}$  are the on-mass-shell
LF momenta, and
\be
        && p^+_1=x_1 P^+, \quad p^+_2=x_2 P^+, \quad x_1+x_2=1\,,\nn \\
        && p_{1\bot}=x_1 P_\bot+k_\bot, \quad p_{2\bot}=x_2
        P_\bot-k_\bot\,,
\label{Pfraction}
\ee
with $(x,k_\perp)$ being the longitudinal momentum fractions 
and transverse momenta of the constituent particles. 
By the Melosh transformation~\cite{Melosh:1974cu},
it is more convenient to work with the following
representation of the wave function:
\be
\Psi^{SS_{z}}(p_{1},p_{2},\lambda_{1},\lambda_{2})=
\frac{1}{\sqrt{2(p_{1}\cdot P+m_{1}M_{0})}}\bar{u}(p_{1},\lambda_{1})
\Gamma_{l,m} u(P,S_{z})\phi(x,k_{\perp})\,,
\label{1/2}
\ee
where $\Gamma_{l,m}$ is the coupling vertex function of the decaying quark $q_{1}$
and the diquark in the baryon state. 
For the scalar diquark, the coupling vertex is $\Gamma_{l,m}=1$.
For the distribution amplitude function of $\phi(x,k_{\perp})$ in Eq.~(\ref{1/2}), 
we use the Gaussian-type function in this work, given by
\be
\phi(x,k_{\perp})=4\left(\frac{\pi}{\beta^{2}}\right)^{3/4}\sqrt{\frac{dk_{z}}{dx}}\exp
\left(\frac{-\vec{k}^{2}}{2\beta^{2}}\right)\,,
\label{DA}
\ee
where $\beta$ is the baryon shape parameter and $k_z$ is defined by
\be
k_z=\frac{xM_0}{2}-\frac{m^2_{2}+k^2_{\perp}}{2xM_0}\,,~~~~~
M_0^2&=&{ m_{1}^2+k_\bot^2\over 1-x}+{ m_{2}^2+k_\bot^2\over  x}\,.
\ee
Using the bound states of
$|\Lambda(P,S,S_z)\rangle$ and $|p (P^{\prime},S^{\prime},S_z^{\prime})\rangle$
in Eq.~(\ref{boundstate}) and the above identities,
we derive the matrix elements of the baryonic transition
in the LF frame. By considering the $\mu=+$ component,
the transition matrix elements are given by
\be
&&\langle {\cal B}_{p}(P^{\prime},S^{\prime},S_z^{\prime})|
\bar{q^{\prime}}\gamma^{+}(1-\gamma_{5}) q|{\cal B}_{\Lambda}(P,S,S_z)\rangle\nonumber \\
&=&
N_{fs}\int{\{d^{4}p_{2}\}}
\frac{\chi_{B_{\Lambda}}(x,{\bf k}_{\perp}) I^{+}
\chi^{\prime}_{B_{p}}(x',{\bf k'}_{\perp})}{(p_{1}^{2}-m_{1}^{2}+i\epsilon)
(p_{1}^{'2}-m_{1}^{'2}+i\epsilon)}\,,
\label{matrix}
\ee
where $I^{+}=\sum_{\lambda_{2}}\bar{u}(\bar{P}',S'_{z})
\left[\bar{\Gamma}^{\prime}_{S(A)}(\strich p_{1}^{\prime}+m_{1}^{\prime})
\gamma^{+}(1-\gamma_{5})(\strich p_{1}+m_{1})\Gamma_{S(A)}\right]u(\bar{P},S_{z})$,
$\bar \Gamma=\gamma^0 \Gamma^\dagger\gamma^0$, 
$\chi_{B_{\Lambda}}$($\chi^{\prime}_{B_{p}})$ corresponds to the vertex function of the 
baryon $\Lambda(p)$,
and 
the flavor spin factor $N_{fs}$ is the overlap factor of particle transformations, 
given by the specific process. 
We refer to the literature~\cite{LFcal1,LFcal2,LFcal3,LFcal4} 
to compute $f_3(g_3)$, consider the case with ``$\mu = +$" and $q^{+} \neq 0$, 
and derive the following relationships:
\be
\bar{u}(P^{\prime},S^{\prime}_{z})q^{+}u(P,S_{z})
&=&\sqrt{P^{+}P^{\prime}}\left[\left(\frac{M_{\Lambda}}{P^{+}}+\frac{M_{p}}{P^{\prime +}}\right)
\delta_{S^{\prime}_{z}S_{z}}+\left(\frac{P^{\prime}_{\perp}}{P^{\prime +}}-
\frac{P_{\perp}}{P^{+}}\right)\sigma^{3}\delta_{S^{\prime}_{z},-S_{z}}
\right]q^{+}
\nonumber \\
\bar{u}(P^{\prime},S^{\prime}_{z})q^{+}\gamma_{5}\,u(P,S_{z})
&=&-\sqrt{P^{+}P^{\prime}}\left[\left(\frac{M_{\Lambda}}{P^{+}}-\frac{M_{p}}{P^{\prime +}}\right)
\sigma^{3}\delta_{S^{\prime}_{z}S_{z}}-\left(\frac{P^{\prime}_{\perp}}{P^{\prime +}}-
\frac{P_{\perp}}{P^{+}}\right)\delta_{S^{\prime}_{z},-S_{z}}
\right]q^{+}\,.
\label{LFspinor}
\ee
Subsequently, in the LF framework, Eq.~(\ref{transitionVA}) can be rewritten as
\be
&&\langle {\cal B}_{p}(P^{\prime},S^{\prime},S_z^{\prime})|
V^{+}|{\cal B}_{\Lambda}(P,S,S_z)\rangle
=-2\sqrt{P^{+}P^{\prime}}\bigg\{f_{1}(q^{2})\delta_{S^{\prime}_{z}S_{z}}
+\frac{f_{2} (q^{2})}{M_{\Lambda}}(\sigma \cdot q_{\perp})\sigma^{3}\,\delta_{S^{\prime}_{z}S_{z}}
\nonumber \\
&&+\frac{f_{3}(q^{2})}{2M_{\Lambda}} 
\left[\left(\frac{M_{\Lambda}}{P^{+}}+\frac{M_{p}}{P^{\prime +}}\right)
\delta_{S^{\prime}_{z}S_{z}}+\left(\frac{P^{\prime}_{\perp}}{P^{\prime +}}-
\frac{P_{\perp}}{P^{+}}\right)\sigma^{3}\delta_{S^{\prime}_{z},-S_{z}}
\right]q^{+}\bigg\}
\nonumber \\
&&\langle {\cal B}_{p}(P^{\prime},S^{\prime},S_z^{\prime})|
A^{+}|{\cal B}_{\Lambda}(P,S,S_z)\rangle
=2\sqrt{P^{+}P^{\prime}}\bigg\{g_{1}(q^{2})\sigma^{3}\delta_{S^{\prime}_{z}S_{z}}
+\frac{g_{2}(q^{2})}{M_{\Lambda}}(\sigma \cdot q_{\perp})\delta_{S^{\prime}_{z}S_{z}}
\nonumber \\
&&-\frac{g_{3}(q^{2})}{2M_{\Lambda}} 
\left[\left(\frac{M_{\Lambda}}{P^{+}}-\frac{M_{p}}{P^{\prime +}}\right)
\sigma^{3}\delta_{S^{\prime}_{z}S_{z}}-\left(\frac{P^{\prime}_{\perp}}{P^{\prime +}}-
\frac{P_{\perp}}{P^{+}}\right)\delta_{S^{\prime}_{z},-S_{z}}
\right]q^{+}\bigg\}\,.
\label{LFff}
\ee
Using the orthogonality relations among the matrices $\delta_{S^{\prime}_{z}S_{z}},\,
\sigma^{3}\delta_{S^{\prime}_{z}S_{z}},\,
\sigma^{i}_{\perp}\sigma^{3}\delta_{S^{\prime}_{z}S_{z}}$, and 
$\sigma^{i}_{\perp}\delta_{S^{\prime}_{z}S_{z}}$, 
the form factors can be extracted through trace operations. 
We note that $f_{1}(q^{2})(g_{1}(q^{2}))$ and $f_{3}(q^{2})(g_{3}(q^{2}))$ 
are orthogonal to $f_{2}(q^{2})(g_{2}(q^{2}))$, which can be derived by using the following 
identities:

\be
\frac{1}{2}\sum_{S^{\prime}_{z}S_{z}}u(P,S_{z})\delta_{S^{\prime}_{z}S_{z}}
\bar{u}(P^{\prime},S^{\prime}_{z})
&=&\frac{1}{4\sqrt{P^{+}P^{\prime}}}(\slashed{P}+M_{0})
\gamma^{+}(\slashed{P}^{\prime}+M^{\prime}_{0})\nonumber\\
\frac{1}{2}\sum_{S^{\prime}_{z}S_{z}}u(P,S_{z})(\sigma^{3})_{S^{\prime}_{z}S_{z}}
\bar{u}(P^{\prime},S^{\prime}_{z})
&=&\frac{1}{4\sqrt{P^{+}P^{\prime}}}(\slashed{P}+M_{0})
\gamma^{+}\gamma_{5}(\slashed{P}^{\prime}+M^{\prime}_{0})\nonumber\\
\frac{1}{2}\sum_{S^{\prime}_{z}S_{z}}u(P,S_{z})(\sigma^{3}\sigma^{i}_{\perp})_{S^{\prime}_{z}S_{z}}
\bar{u}(P^{\prime},S^{\prime}_{z})
&=&\frac{-i}{4\sqrt{P^{+}P^{\prime}}}(\slashed{P}+M_{0})
\sigma^{i+}(\slashed{P}^{\prime}+M^{\prime}_{0})\nonumber\\
\frac{1}{2}\sum_{S^{\prime}_{z}S_{z}}u(P,S_{z})(\sigma^{i}_{\perp})_{S^{\prime}_{z}S_{z}}
\bar{u}(P^{\prime},S^{\prime}_{z})
&=&\frac{i}{4\sqrt{P^{+}P^{\prime}}}(\slashed{P}+M_{0})
\sigma^{i+}\gamma_{5}(\slashed{P}^{\prime}+M^{\prime}_{0})
\label{identity}
\ee
Through Eqs.~(\ref{matrix}),~(\ref{LFff}), and~(\ref{identity}), 
we obtain the expressions of the form factors as follows:
\be
&&f_{1}(q^{2})\delta_{S^{\prime}_{z}S_{z}}
+\frac{f_{3}(q^{2})}{2M_{\Lambda}} \left(\frac{M_{\Lambda}}{P^{+}}
+\frac{M_{p}}{P^{\prime +}}\right)
q^{+}\,\delta_{S^{\prime}_{z}S_{z}} \nonumber \\
&=&\frac{N_{fs}}{8P^{+}P^{\prime +}}\int{\{d^{4}p_{2}\}}
\frac{\chi_{B_{\Lambda}}(x,{\bf k}_{\perp}) 
\chi^{\prime}_{B_{p}}(x',{\bf k'}_{\perp})}{(p_{1}^{2}-m_{1}^{2}+i\epsilon)
(p_{1}^{'2}-m_{1}^{'2}+i\epsilon)}\,\nonumber \\
&&\times {\rm Tr} [ (\slashed P+M_{0})\gamma^{+}(\slashed P^{\prime}+M_{0}^{\prime})
(\slashed p_{1}^{\prime}+m_{1}^{\prime})\gamma^{+}
(\slashed p_{1}+m_{1}) ]\,,
\label{f3}
\ee
\be
&&g_{1}(q^{2})\sigma^{3}\delta_{S^{\prime}_{z}S_{z}}
-\frac{g_{3}(q^{2})}{2M_{\Lambda}} \left(\frac{M_{\Lambda}}{P^{+}}
-\frac{M_{p}}{P^{\prime +}}\right)
\,\sigma^{3}q^{+}\,\delta_{S^{\prime}_{z}S_{z}}\nonumber \\
&=& \frac{N_{fs}}{8P^{+}P^{\prime +}}\int{\{d^{4}p_{2}\}}
\frac{\chi_{B_{\Lambda}}(x,{\bf k}_{\perp}) 
\chi^{\prime}_{B_{p}}(x',{\bf k'}_{\perp})}{(p_{1}^{2}-m_{1}^{2}+i\epsilon)
(p_{1}^{'2}-m_{1}^{'2}+i\epsilon)}\,\nonumber \\
&&\times  
{\rm Tr}[ (\slashed P+M_{0})\gamma^{+}\gamma_{5}(\slashed P^{\prime}+M_{0}^{\prime})
(\slashed p_{1}^{\prime}+m_{1}^{\prime})\gamma^{+}\gamma_{5}
(\slashed p_{1}+m_{1}) ]\,.
\label{g3}
\ee
\be
&&\frac{f_{2} (q^{2})}{M_{\Lambda}}(\sigma \cdot q_{\perp} \sigma^{3})\delta_{S^{\prime}_{z}S_{z}} \nonumber \\
&=& \frac{N_{fs}}{8P^{+}P^{\prime +}}\int{\{d^{4}p_{2}\}}
\frac{\chi_{B_{\Lambda}}(x,{\bf k}_{\perp}) 
\chi^{\prime}_{B_{p}}(x',{\bf k'}_{\perp})}{(p_{1}^{2}-m_{1}^{2}+i\epsilon)
(p_{1}^{'2}-m_{1}^{'2}+i\epsilon)}\,\nonumber \\
&&\times{\rm Tr} [ (\slashed P+M_{0})\sigma^{\nu +}(\slashed P^{\prime}+M_{0}^{\prime})
(\slashed p_{1}^{\prime}+m_{1}^{\prime})\gamma^{+}
(\slashed p_{1}+m_{1}) ]\,\,, \nonumber \\
&&\frac{g_{2}(q^{2})}{M_{\Lambda}}(\sigma \cdot q_{\perp})\delta_{S^{\prime}_{z}S_{z}}
\nonumber \\
&=& \frac{N_{fs}}{8P^{+}P^{\prime +}}\int{\{d^{4}p_{2}\}}
\frac{\chi_{B_{\Lambda}}(x,{\bf k}_{\perp}) 
\chi^{\prime}_{B_{p}}(x',{\bf k'}_{\perp})}{(p_{1}^{2}-m_{1}^{2}+i\epsilon)
(p_{1}^{'2}-m_{1}^{'2}+i\epsilon)}\,\nonumber \\
&&\times{\rm Tr} [ (\slashed P+M_{0})\sigma^{\nu +}\gamma_{5}(\slashed P^{\prime}+M_{0}^{\prime})
(\slashed p_{1}^{\prime}+m_{1}^{\prime})\gamma^{+}\gamma_{5}
(\slashed p_{1}+m_{1}) ]\,\,.
\label{f2g2}
\ee
where $\nu = 1, 2$ and the variables of $(x',{\bf k'}_\perp)$ 
are the LF relative momentum 
variables of ${\cal B}_{f}(P^{\prime},S^{\prime},S_z^{\prime})$
with the definitions given by replacing $x\to x'$ in Eq.~(\ref{Pfraction}). 
The extraction of $f_2$ and $g_2$ is performed before taking the $q_{\perp}\to 0$ limit.
Furthermore, since the system is analyzed in the timelike domain 
(i.e., $q^2 = q^{+}q^{-} - q^2_{\perp} \geq 0$), momentum is transmitted only 
in the longitudinal direction, i.e., $q_{\perp} = 0$. 
Therefore, we can choose a reference frame such that $P(P^{\prime})_{\perp} = 0$ 
which causes the term containing $\delta_{S^{\prime}_{z},-S_{z}}$ 
in Eqs.~(\ref{LFspinor}) and (\ref{LFff}) to vanish. 
To decouple $f_{1}(g_{1})$ and $f_{3}(g_{3})$, we rewrite Eqs.~(\ref{f3}) and (\ref{g3}) as
\be
f_{1}(q^{2}) + A f_{3}(q^{2}) = Hv(q^{2})
&&=\frac{N_{fs}}{8P^{+}P^{\prime +}}\int{\{d^{4}p_{2}\}}
\frac{\chi_{B_{\Lambda}}(x,{\bf k}_{\perp}) 
\chi^{\prime}_{B_{p}}(x',{\bf k'}_{\perp})}{(p_{1}^{2}-m_{1}^{2}+i\epsilon)
(p_{1}^{'2}-m_{1}^{'2}+i\epsilon)}\,\nonumber \\
&&\times 
{\rm Tr} [ (\slashed P+M_{0})\gamma^{+}(\slashed P^{\prime}+M_{0}^{\prime})
(\slashed p_{1}^{\prime}+m_{1}^{\prime})\gamma^{+}
(\slashed p_{1}+m_{1}) ] \,,\nonumber \\
g_{1}(q^{2}) + B g_{3}(q^{2}) = Ha(q^{2})
&&= \frac{N_{fs}}{8P^{+}P^{\prime +}}\int{\{d^{4}p_{2}\}}
\frac{\chi_{B_{\Lambda}}(x,{\bf k}_{\perp}) 
\chi^{\prime}_{B_{p}}(x',{\bf k'}_{\perp})}{(p_{1}^{2}-m_{1}^{2}+i\epsilon)
(p_{1}^{'2}-m_{1}^{'2}+i\epsilon)}\,\nonumber \\
&&\times  
{\rm Tr}[ (\slashed P+M_{0})\gamma^{+}\gamma_{5}(\slashed P^{\prime}+M_{0}^{\prime})
(\slashed p_{1}^{\prime}+m_{1}^{\prime})\gamma^{+}\gamma_{5}
(\slashed p_{1}+m_{1}) ] \,,
\label{f3g3}
\ee
where
\be
A= \frac{1}{2M_{\Lambda}}\left(\frac{M_{\Lambda}}{P^{+}}
+\frac{M_{p} }{P^{\prime +}}\right)q^{+}
=\frac{1}{2M_{\Lambda}}\frac{1-\alpha}{\alpha}(\alpha M_{\Lambda}+M_{p})\nonumber \\
B= -\frac{1}{2M_{\Lambda}}\left(\frac{M_{\Lambda} }{P^{+}}
-\frac{M_{p} }{P^{\prime +}}\right)q^{+}
=-\frac{1}{2M_{\Lambda}}\frac{1-\alpha}{\alpha}(\alpha M_{\Lambda}-M_{p})
\label{AB}
\ee
where the minus sign in the axial contribution has been absorbed into the definition of $B$ 
and $\alpha=P^{\prime+} /P^{+}$. There are two solutions for $\alpha$, namely, 
\be
\alpha_{\pm}
=\frac{M^{2}_{\Lambda}+M^{2}_{p}-q^{2}\pm\sqrt{(M^{2}_{\Lambda}+M^{2}_{p}-q^{2})^{2}
-4 M^{2}_{\Lambda} M^{2}_{p}} }{2 M^{2}_{\Lambda}}\,.
\label{eq21}
\ee
Based on the definition of Eq.~(\ref{eq21}), 
the $+(-)$ signs correspond to the outgoing baryon recoiling 
in the positive (negative) $z$ direction relative to incoming baryon. The form factors are 
independent of the reference frame chosen for the moving direction. 
Therefore, $\alpha$ can be decomposed into two parts, 
$\alpha_{+}$ and $\alpha_{-}$, to solve for $f_{1}(g_{1})$ and $f_{3}(g_{3})$  
where the consistency between $\alpha_{+}$ and $\alpha_{-}$ 
serve as a self-consistency check:
\be
f_{1}(q^{2}) + A_{+} f_{3}(q^{2}) = Hv(q^{2})|_{\alpha=\alpha_{+}}\,,
\nonumber \\
f_{1}(q^{2}) + A_{-} f_{3}(q^{2}) = Hv(q^{2})|_{\alpha=\alpha_{-}}\,,
\label{f3g3-2}
\ee
where $A_{+}=A|_{\alpha=\alpha_{+}}$ and $A_{-}=A|_{\alpha=\alpha_{-}}$. 
One obtains
\be
f_{1}(q^2)&=&\frac{A_{-} Hv(q^{2})|_{\alpha=\alpha_{+}}
-A_{+} Hv(q^{2})|_{\alpha=\alpha_{-}}}{A_{-}-A_{+}},\nonumber \\
f_{3}(q^2)&=&\frac{Hv(q^{2})|_{\alpha=\alpha_{-}}
-Hv(q^{2})|_{\alpha=\alpha_{+}}}{A_{-}-A_{+}}.
\ee
By repeating the steps from Eqs.~(\ref{f3g3}) to (\ref{f3g3-2}), we obtain
\be
g_{1}(q^2)&=&\frac{B_{-} Ha(q^{2})|_{\alpha=\alpha_{+}}
-B_{+} Ha(q^{2})|_{\alpha=\alpha_{-}}}{B_{-}-B_{+}},\nonumber \\
g_{3}(q^2)&=&\frac{Ha(q^{2})|_{\alpha=\alpha_{-}}
-Ha(q^{2})|_{\alpha=\alpha_{+}}}{B_{-}-B_{+}}.
\ee
where $B_{+}=B|_{\alpha=\alpha_{+}}$ and $B_{-}=B|_{\alpha=\alpha_{-}}$.

\begin{figure}[h]
\includegraphics{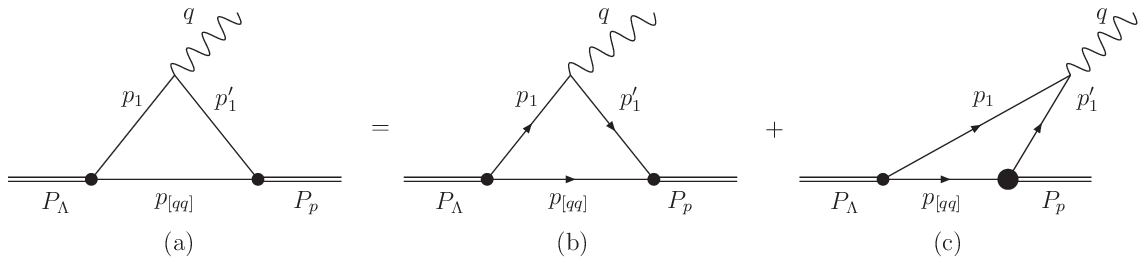}
\vskip 4cm
\caption{The effective treatment of the LF amplitude (a) can be decomposed into
the LF valence part (b) in $0 < x <\alpha$ and the nonvalence one (c)
in $\alpha < x < 1$, where
the small and large solid circles of the mediator-quark vertices 
in (b) and (c) represent the LF ordinary
and nonvalence wave function vertices, respectively.}
\end{figure}

\sse{Valence contribution}

The trace term $I^{+}$ in Eq.~(\ref{matrix}) can be written as
the sum of the valence $I^{+}_{V}$ and nonvalence 
$I^{+}_{NV}$ parts, as shown in Fig.~1(a).
In the region  of $0<x<\alpha$ with $0<p^{+}< P'^{+}$ and $p^{-}_{[qq]}=p^{-}_{[qq]on}=
(m_{[qq]}^{2}+k^{2}_{\perp})/p^{+}_{[qq]}$, in which the subscript ``$on$"
indicates on the mass shell, called the valence region as seen in Fig.~1(b),
the effective contribution of the LF valence amplitude is given by
\be
&&{\cal M}_{val}
=\frac{N_{fs}}{16\pi^3}
\int^{\alpha}_{0}dx\int d^{2}{\bf k}_{\perp}
\Phi_{_{\Lambda}}(x,{\bf k}_{\perp}) I^{+}_{V}
\Phi_{p}(x',{\bf k'}_{\perp}),\nonumber \\
&&I^{+}_{V}=\bar{u}(\bar{P}^{\prime},S_{z}^{\prime})\bar{\Gamma}^{\prime}
(\strich p_{1}^{\prime}+m_{1}^{\prime})\gamma^{+}(1-\gamma_{5})
(\strich p_{1}+m_{1})\Gamma u(\bar{P},S_{z})\,.
\label{ValM}
\ee
Following Refs.~\cite{LFcal1,LFcal2,LFcal3,LFcal4}, we drive the transition form factors:
\be
Hv_{val}(q^{2})&=&\frac{N_{fs}}{16\pi^3}
\int^{\alpha}_{0}dx\int d^{2}{\bf k}_{\perp}
\Phi_{\Lambda}(x,{\bf k}_{\perp}) \Phi_{p}(x',{\bf k'}_{\perp})
\nonumber \\
&&\times[k_{\perp}\cdot k_{\perp}^{\prime}+((1-x) M_{0}+m_{1})
((1-x^{\prime})M_{0}^{\prime}+m_{1}^{\prime})]\,,\nonumber \\
Ha_{val}(q^{2})&=& \frac{N_{fs}}{16\pi^3}
\int^{\alpha}_{0}dx\int d^{2}{\bf k}_{\perp}
\Phi_{\Lambda}(x,{\bf k}_{\perp}) \Phi_{p}(x',{\bf k'}_{\perp})
\nonumber \\
&&\times[-k_{\perp}\cdot k_{\perp}^{\prime}+((1-x) M_{0}+m_{1})
((1-x^{\prime})M_{0}^{\prime}+m_{1}^{\prime})]\,,
\label{Hva}
\ee
\be
\frac{f_{2_{val}}(q^{2})}{M}&=& \frac{N_{fs}}{16\pi^3 q_{\perp}^{2}}
\int^{\alpha}_{0}dx\int d^{2}{\bf k}_{\perp}
\Phi_{\Lambda}(x,{\bf k}_{\perp})
\Phi_{p}(x',{\bf k'}_{\perp})\nonumber \\
&&\times[-(m_{1}+(1-x) M_{0})k_{\perp}^{\prime}\cdot q_{\perp}
+(m_{1}^{\prime}+(1-x^{\prime})M_{0}^{\prime})k_{\perp}\cdot q_{\perp}]\,,\nonumber \\
\frac{g_{2_{val}}(q^{2})}{M}&=& \frac{N_{fs}}{16\pi^3 q_{\perp}^{2}}
\int^{\alpha}_{0}dx\int d^{2}{\bf k}_{\perp}
\Phi_{\Lambda}(x,{\bf k}_{\perp})
\Phi_{p}(x',{\bf k'}_{\perp})\nonumber \\
&&\times[-(m_{1}+(1-x) M_{0})k_{\perp}^{\prime}\cdot q_{\perp}
-(m_{1}^{\prime}+(1-x^{\prime})M_{0}^{\prime})k_{\perp}\cdot q_{\perp}]
\label{f2g2val}
\ee
with
\be
\Phi_{B}(x,{\bf k}_{\perp})=\frac{\phi(x,k_{\perp})}
{\sqrt{2x_{1}(p_{1}\cdot P+m_{1}M_{0})}}
\ee
\sse{Nonvalence contribution}

In the nonvalence region of $\alpha<x<1$, $P'^{+}<p^{+}<P^{+}$, and $p_{1}^{-}=p^{-}_{1on}=
(m_{1}^{2}+k^{2}_{1\perp})/p^{+}_{1}$,
as shown in Fig.~1(c),
the trace term in Eq.~(\ref{matrix}) can be separated into
the on-shell propagating and instantaneous parts of $I^{\mu}_{on}$ and $I^{\mu}_{inst}$ via
\be
\slashed p+m=(\slashed p_{on}+m)+\frac{1}{2}\gamma^{+}(p^{-}-p^{-}_{on})\,,
\label{eq24}
\ee
respectively. The effective contribution can be found as
\be
{\cal M}_{non-val}&=&\frac{N_{fs}}{16\pi^3}
\int^{1}_{\alpha}dx
\int d^{2}{\bf k}_{\perp}\Gamma_g(x,{\bf k}_{\perp})I^{+}_{NV}
\Phi_{\Lambda}(x,{\bf k}_{\perp})\nonumber\\
&\times&
\int \frac{dy}{y(1-y)}\int d^2{\bf l}_{\perp}
{\cal K}(x,{\bf k}_{\perp};y,{\bf l}_{\perp})
\Phi_{p}(y,{\bf l}_{\perp})\,,
\label{NVmatrix}
\ee
where $I^{+}_{NV}$ is the trace term in the nonvalence region.
Substituting Eq.~(\ref{eq24}) into $I^{+}$, we  get 
$I^{+}_{NV}=I^{+}_{V}(p_{i}^{-}=p^{-}_{ion}=(m_{i}^{2}+k^{2}_{i\perp})/p^{+}_{i})
+I^{+}_{inst}$.
The LF vertex function of a gauge boson
$\Gamma_g$ in Eq.~(\ref{NVmatrix}) corresponds to the LF
energy denominator with its explicit form  given by~\cite{BS1,BS3,LFg}
\be
\Gamma_g^{-1}(x,{\bf k}_\perp)=
\alpha\biggl[\frac{q^2}{1-\alpha} -
\biggl(\frac{{\bf k}^2_\perp + m^2_1}{1-x}
+\frac{{\bf k'}^2_\perp + m'^{2}_{1}}{\alpha-x}\biggr)
\biggr].
\label{gaugeWF}
\ee
The trace terms in Eqs.~(\ref{ValM}) and (\ref{NVmatrix}),
both corresponding to the products of the initial and final LF spin wave functions,
can be obtained by off-shell Melosh transformations.
The form factors related to the nonvalence diagram $I^+_{NV}$ are given by
\be
Hv_{NV}(q^{2})&=&\frac{N_{fs}}{16\pi^3}
\int^{1}_{\alpha}dx
\int d^{2}{\bf k}_{\perp}\Gamma_g(x,{\bf k}_{\perp})\Phi_{\Lambda}(x,{\bf k}_{\perp}) \nonumber \\
&\times&\{[k_{\perp}\cdot k_{\perp}^{\prime}+((1-x)M_{0}+m_{1})
((1-x^{\prime})M_{p}+m_{1}^{\prime})]+I^{+}_{inst}\}\nonumber\\
&\times&
\int \frac{dy}{y(1-y)}\int d^2{\bf l}_{\perp}
{\cal K}(x,{\bf k}_{\perp};y,{\bf l}_{\perp})
\Phi_{p}(y,{\bf l}_{\perp})\,, \nonumber\\
Ha_{NV}(q^{2})&=& \frac{N_{fs}}{16\pi^3}
\int^{1}_{\alpha}dx
\int d^{2}{\bf k}_{\perp}\Gamma_g(x,{\bf k}_{\perp})\Phi_{\Lambda}(x,{\bf k}_{\perp})\nonumber\\
&\times&\{[-k_{\perp}\cdot k_{\perp}^{\prime}+((1-x)M_{0}+m_{1})
((1-x^{\prime})M_{p}+m_{1}^{\prime})]+I^{+}_{inst}\}\nonumber\\
&\times&
\int \frac{dy}{y(1-y)}\int d^2{\bf l}_{\perp}
{\cal K}(x,{\bf k}_{\perp};y,{\bf l}_{\perp})
\Phi_{p}(y,{\bf l}_{\perp})\,,
\label{f1g1NV}
\ee
\be
\frac{f_{2_{NV}}(q^{2})}{M}&=& \frac{N_{fs}}{16\pi^3 q_{\perp}^{2}}
\int^{1}_{\alpha}dx
\int d^{2}{\bf k}_{\perp}\Gamma_g(x,{\bf k}_{\perp})\Phi_{\Lambda}(x,{\bf k}_{\perp})\nonumber\\
&\times&\{
[(m_{1}+(1-x)M_{0})k_{\perp}^{\prime}\cdot q_{\perp}
-(m_{1}^{\prime}+(1-x^{\prime})M_{p})k_{\perp}\cdot q_{\perp}]+I^{\prime +}_{inst}\}
\nonumber\\
&\times&
\int \frac{dy}{y(1-y)}\int d^2{\bf l}_{\perp}
{\cal K}(x,{\bf k}_{\perp};y,{\bf l}_{\perp})
\Phi_{p}(y,{\bf l}_{\perp})\,,\nonumber\\
\frac{g_{2_{NV}}(q^{2})}{M}&=& \frac{N_{fs}}{16\pi^3 q_{\perp}^{2}}
\int^{1}_{\alpha}dx
\int d^{2}{\bf k}_{\perp}\Gamma_g(x,{\bf k}_{\perp})\Phi_{\Lambda}(x,{\bf k}_{\perp})\nonumber\\
&\times&\{
[-(m_{1}+(1-x)M_{0})k_{\perp}^{\prime}\cdot q_{\perp}
-(m_{1}^{\prime}+(1-x^{\prime})M_{p})k_{\perp}\cdot q_{\perp}]+I^{\prime +}_{inst}\}
\nonumber\\
&\times& \int \frac{dy}{y(1-y)}\int d^2{\bf l}_{\perp} 
{\cal K}(x,{\bf k}_{\perp};y,{\bf l}_{\perp}) \Phi_{p}(y,{\bf
l}_{\perp})\,, \label{f2g2NV} 
\ee 
where $I^{+}_{inst}=I^{\prime +}_{inst}=0$. 
The complete form factors are $f(g)_{j}=f(g)_{j_{V}}+f(g)_{j_{NV}}$ ($j=1,2$).
Notice that the instantaneous contribution exists
in the nonvalence diagram only when the ``$+$" component is used.
The vertex of the nonvalence wave function is usually obtainable
from the Bethe-Salpeter (B-S) amplitude in the B-S theory~\cite{FVNV,BS1}.
The corresponding light-front bound-state equation can be written as~\cite{BS1,BSEQ2,BSEQ3}
\be
(M^{2}-M^{2}_{0})\Phi'(x_{i},{k}_{i\perp})
=\int [dy][d^2{\bf l}_{\perp}]
{\cal K}(x_{i},{\bf k}_{i\perp};y,{\bf l}_{\perp})
\Phi(y,{\bf l}_{\perp})\,.
\label{LFBS}
\ee
Both valence and nonvalence B-S amplitudes can be regarded as solutions of Eq.~(\ref{LFBS}).
The normal and nonvalence B-S amplitudes correspond 
to  $x < \alpha$ and  $x > \alpha$, respectively.
In Fig.~1(c), the nonvalence B-S amplitude can be phenomenologically effectively related through 
an analytic continuation procedure of the valence B-S amplitude.
In the LFQM, the relationship between the B-S amplitudes of
two regions is given in Refs.~\cite{BS1,BS2,BS3}.
However, for the integral equation of Eq.~(\ref{LFBS})
it is necessary to use the nonperturbative quantum chromodynamics (QCD) 
method to obtain the kernel.

\begin{figure}[h]
\includegraphics{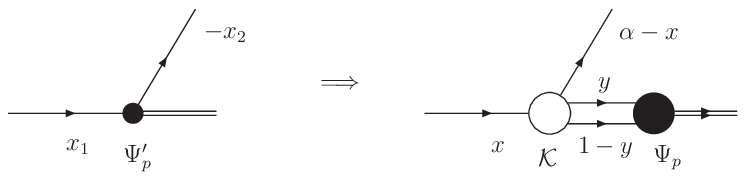}
\vskip 4cm
\caption{Relationship between the ordinary LF 
wave function (large solid vertex) and the nonvalence vertex (small solid vertex)}
\label{figki}
\end{figure}

The relevant operator ${\cal K}$ in Eq.~(\ref{LFBS}) is the B-S core,
which in principle contains contributions from high Fock states. 
As shown in Fig.~\ref{figki}, the kernel ${\cal K}$ provides the 
dynamical connection between the higher-Fock and ordinary valence configurations. 
We define that
\be
G_{B_{\Lambda}B_{p}}\equiv\int[dy][d^2{\bf l}_{\perp}]
{\cal K}(x_{i},{\bf k}_{i\perp};y,{\bf }_\perp)\Phi_{_{p}}(y,{\bf l}_{\perp})\,,
\ee
which depends only on $x$ and ${\bf k}_{\perp}$.
The range of the momentum fraction $x$ relies on the external momenta for the embedded states. 
In this study, we approximate $G_{{\cal B}_{\Lambda}{\cal B}_{p}}$ as a constant 
because the initial wave function $\Psi_{\Lambda}(x, k_{\perp})$ acts as a 
weighting factor in the nonvalence state contribution, 
which causes the initial Gaussian wave function to act as a suppressor, 
with the main contribution originating from regions with narrow $x$ values 
and low transverse momentum $k_{\perp}$. Consequently, 
the momentum dependence of $G_{{\cal B}_{\Lambda}{\cal B}_{p}}$ becomes effectively weak, 
allowing us to approximate it as a constant parameter
~\cite{BS1,BS2,BS3}. 
Although this approximation has mainly been tested in mesonic systems, 
the same suppression mechanism from the Gaussian wave function is 
expected to persist in the quark-diquark description of baryons.

\se{Numerical Results And Discussions}

\sse{Form factors}
To numerically evaluate the exclusive transition form factors in the LFQM,
we work directly in the timelike region.
In the previous calculations, we assumed $q_{\perp}\neq 0$ 
and only set $q_{\perp}= 0$ in the final numerical calculation. 
In the numerical analysis, the quark masses of $m_s$ are taken from Ref.~\cite{LFQM1}; 
$m_u$ is set to be 0.25 in GeV; 
the mass of diquark $m_{[ud]}$, parameters $\beta_{s,[ud]}$ 
and $\beta_{u,[ud]}$ are chosen from Refs.~\cite{diquark2,lf4,LFQM1}. 
The input parameters are 
shown in Table~\ref{parameter}.
\begin{table}[htbp]
\caption{Input parameters for the $\Lambda \to p$ transitions.}
\vskip 0.2in
\label{parameter}
\begin{tabular}{ c c  c  c  c } \hline
$m_{u,d}$ & $m_{s}$ & $m_{[qq']}$ & $~\beta_{u[qq]}$ & $~\beta_{s[qq]}$ 
\\ \hline \hline
$0.25$ & $0.48$ & $~0.4-0.7$ & $~0.34\pm0.02$ & $~0.38\pm0.02$ 
\\ \hline
\end{tabular}
\end{table}
In Eq.~(\ref{NVmatrix}), $G_{{\cal B}_{\Lambda}{\cal B}_{p}}$ is treated as
a constant in the range of $1.0-6.0$, which was previously tested in
some exclusively semileptonic mesonic decay processes and shown
to be a good approximation for processes with a small momentum transfer~\cite{BS1,BS2,BS3}. 
Explicitly, in our numerical evaluation we take 
$G_{{\cal B}_{\Lambda}{\cal B}_{p}}=2$, which  
 is generally determined by comparison with the experimental results. 
Furthermore, we use $N_{fs}$=$\frac{1}{\sqrt{6}}$ with
 detailed calculation given in the Appendix of Ref.~\cite{LFQM1}.

To describe the momentum $q^2$ behaviors,
we parametrize the form factor using the double-pole forms of
\be
F(q^2)=\frac{F(0)}{1+a(q^2/M_{pole}^{2})+b(q^4/M_{pole}^{4})}\,,
\label{fit1}
\ee
where $M_{pole}=1.115$ GeV is treated as an effective hadronic scale characterizing 
the momentum dependence of the form factors, and $(F(0),a,b)$ 
can be determined in the numerical analysis. 
The sign convention in Eq. (\ref{fit1}) follows the fitting definition adopted in the present work.
Another form of parametrizing the form factors is:
\be
{\cal F}(q^{2})=\frac{1}{1-\frac{q^{2}}{M_{pole}}}
\left[F(0)+C\left(z(q^{2})-z(0)\right)\right]\,,~~~~
z(q^{2})=\frac{\sqrt{t_{+}-q^{2}}-\sqrt{t_{+}-t_{0}}}
{\sqrt{t_{+}-q^{2}}+\sqrt{t_{+}-t_{0}}}
\label{fit2}
\ee
where $t_{\pm}=(M_{\Lambda}\pm M_{p})^2$ and $t_{0}=t_{+}(1-\sqrt{t_{-}/t_{+}})$. 
For Eqs.~(\ref{fit1}) and (\ref{fit2}), describing the $q^2$ dependence of the form factors, 
the results of the parametrization are shown in Table~\ref{Table1}. 
From Table~\ref{Table1}, $g_3(0)$ is numerically large. This is mainly because, 
in Eq.~(\ref{AB}), the value of $B$ is mathematically about 
one order of magnitude smaller than that of $A$. 
Physically, $g_3(0)$ contributes only through the scalar helicity amplitude 
and is further suppressed by the lepton mass. 
Additionally, in Eq.~(\ref{fit2}), the value of the fitting parameter associated 
with $g_3(q^2)$ is large because the pseudoscalar contribution exhibits strong momentum 
dependence in the timelike domain. All fitting parameters are derived primarily 
from the momentum dependence of the form factor rather than the physical observables themselves. 
Although the fitted coefficients associated with $g_3$ become numerically large, 
their impact on the physical decay observables remains moderate due to 
the lepton-mass suppression of the corresponding helicity amplitudes.

\begin{figure}[t!]
\includegraphics[width=3.1in]{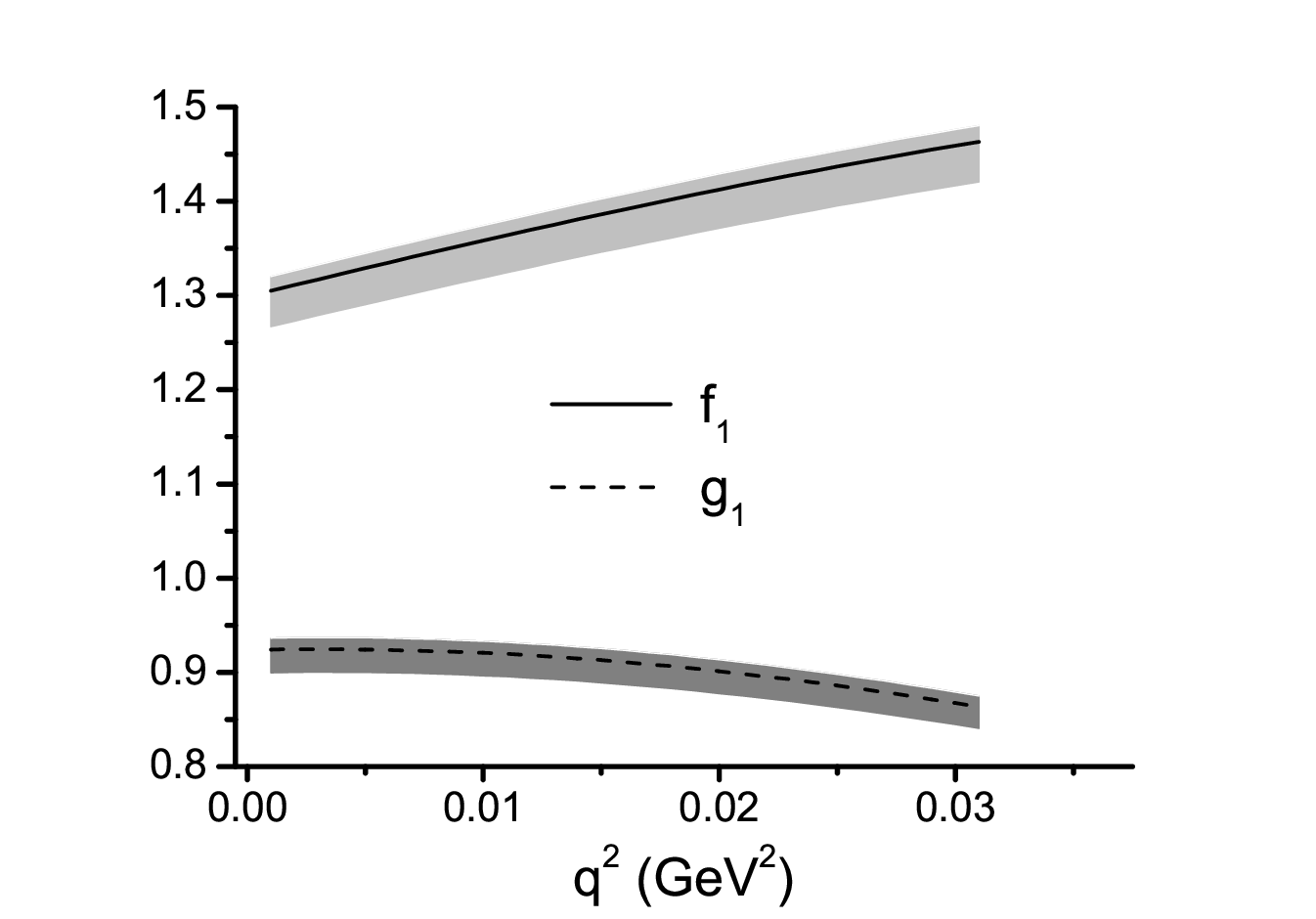}
\includegraphics[width=3.1in]{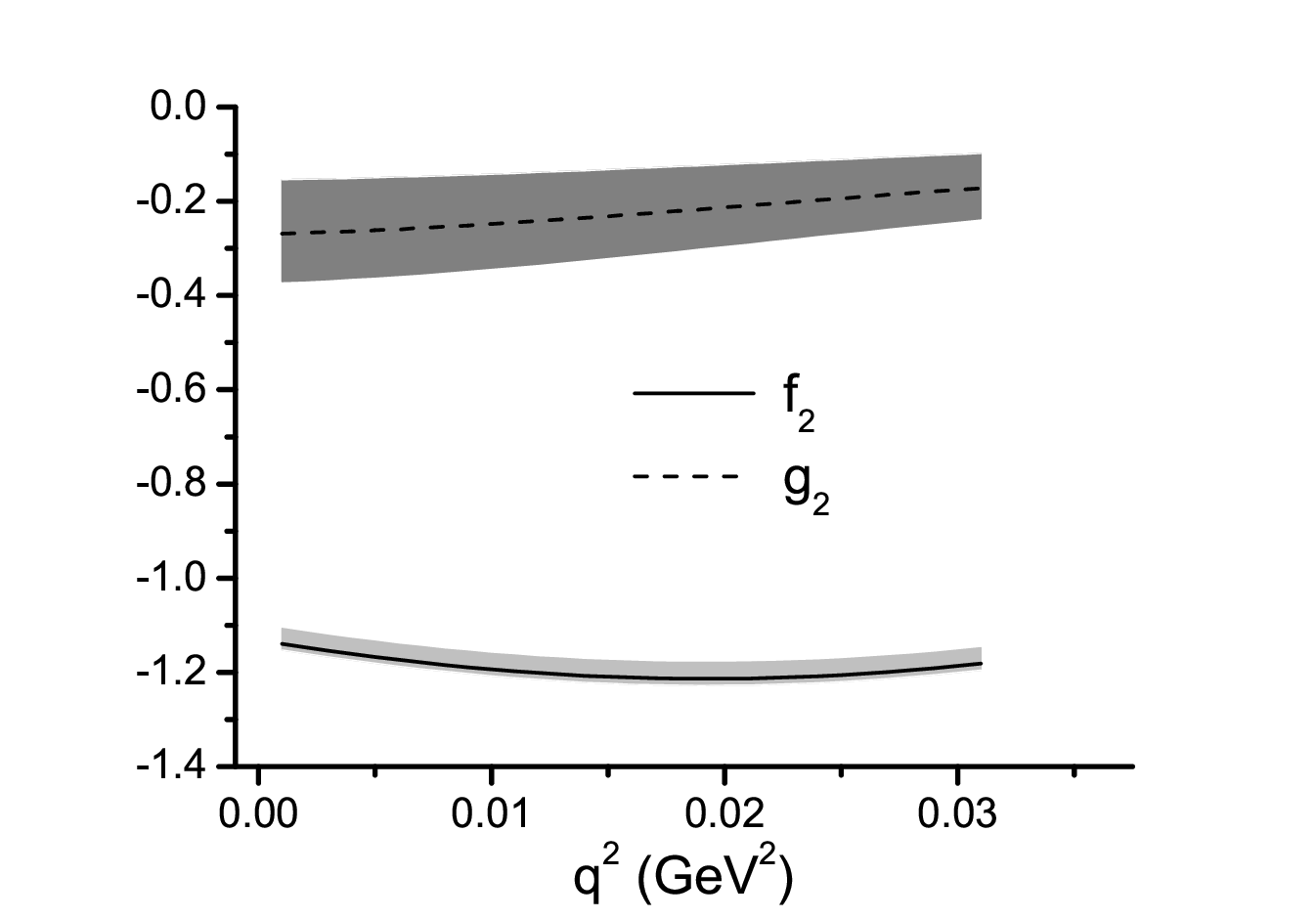}
\includegraphics[width=3.1in]{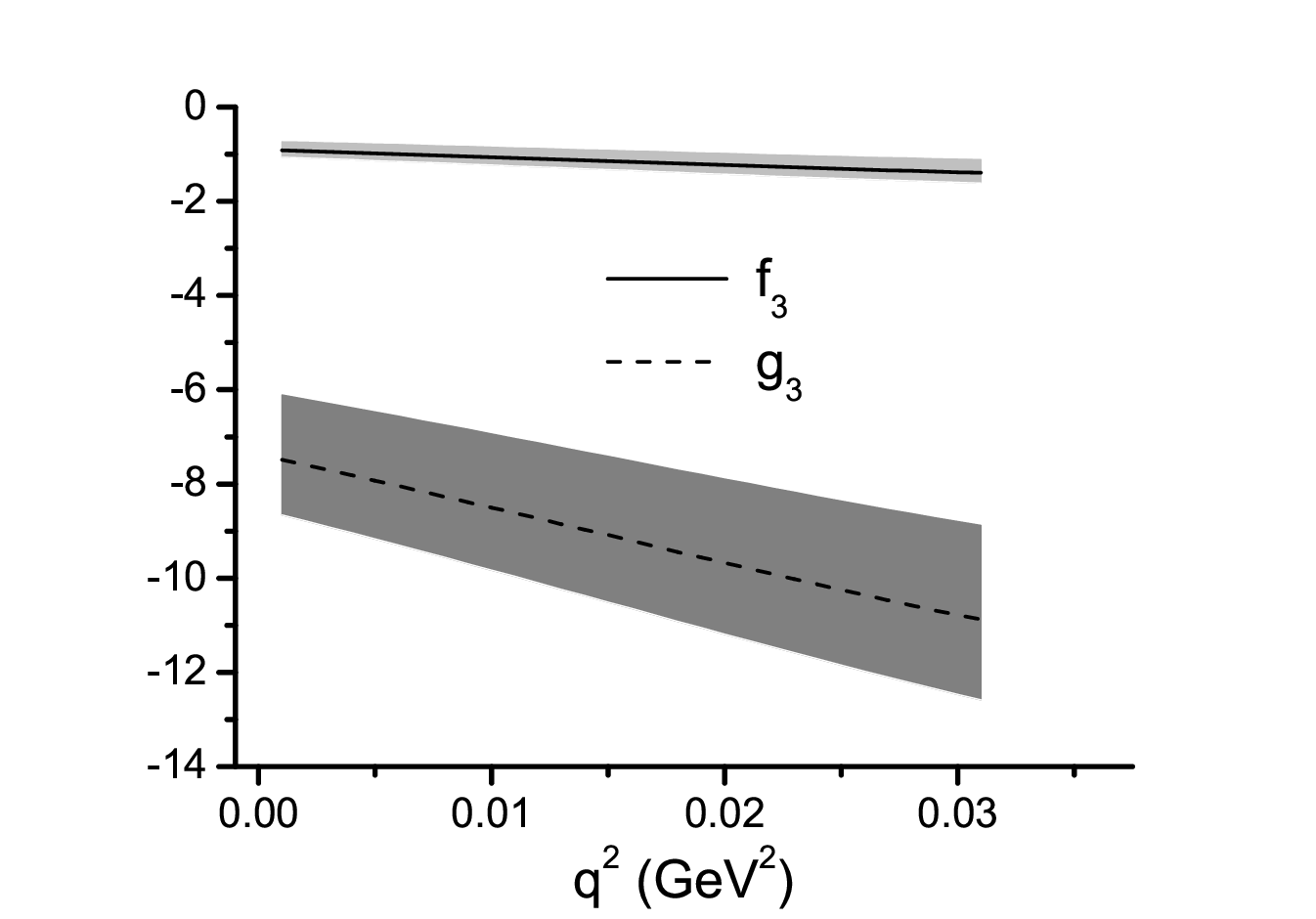}
\caption{Transition form factors for $\Lambda\to p$. 
The gray bands denote the uncertainties associated with the shape parameters.}
\label{fig2}
\end{figure}

\begin{table}[htbp]
\caption{Form factors of the $\Lambda \to p$ transitions.}
\vskip 0.2in
\label{Table1}
\begin{tabular}{|c||c|c|c|c|c|c|} \hline
$\Lambda \to p$ & $f_{1}$ & $f_{2}$ & $f_{3}$ & 
$g_{1}$ & $g_{2}$ & $g_{3}$
\\ \hline \hline
$F(0)$ & $1.305^{+0.18}_{-0.1}$ & $-1.139^{+0.123}_{-0.235}$ & $-0.917^{+0.212}_{-0.450}$ 
& $0.924^{+0.137}_{-0.062}$ & $-0.269^{+0.079}_{-0.022}$ & $-7.487^{+1.537}_{-3.491}$
\\ \hline
$a$ & $-5.976$ & $-8.562$ & $-20.152$ 
& $-0.464$ & $4.975$ & $-17.325$
\\ \hline
$b$ & $55.964$ & $303.263$ & $257.322$
& $143.982$ & $833.088$ & $197.214$
\\ \hline\hline
$C$ & $-80.119$ & $96.730$ & $21.09$ 
& $61.492$ & $8.846$ & $16152.9$ 
\\ \hline
\end{tabular}
\end{table}
In Fig.~\ref{fig2}, we present the numerical results of the form factors as
functions of $q^2$ for the $\Lambda\to p$ transition. 
In Table~\ref{Table2}, we compare the obtained form factors, expressed as constant ratios, 
with experimental and predicted values from several phenomenological models, 
where $f^{SU(3)}_{1}=\sqrt{3/2}$.~\cite{chiral} 
We observe that the results for $g_{1}/f_{1}$ are in good agreement 
with the latest results from the BESIII experiments and QCD sum rule calculations, 
except that $f_{2}/f_{1}$ is slightly larger. 
Overall, the largest deviations from the results of lattice QCD~\cite{Lattic} 
are $f_{2}/f_{1}$ and $g_{2}/f_{1}$, mainly because our results of $f_3$ and $g_2$ 
are significantly larger, especially $g_2$, while the other ratios are basically 
consistent within the error range. 
We also note that the sign of $f_{2}/f_{1}$ differs from that of some existing literature. 
This sign depends on the definition of the form factor, 
and different methods in the literature define $f_{2}$ differently.
\begin{table}[htbp]
\caption{Ratios of form factors from the $\Lambda \to p$ transition.}
\vskip 0.2in
\label{Table2}
\begin{tabular}{|c||c|c|c|c|} \hline
$\Lambda \to p$ & $f_{1}/f_{1}^{SU(3)}$ & $g_{1}/f_{1}$ & $f_{2}/f_{1}$ & 
$g_{2}/f_{1}$ 
\\ \hline \hline
Our results & $1.065^{+0.18}_{-0.1}$ & $0.708^{+0.144}_{-0.072}$ & 
$-0.872^{+0.153}_{-0.192}$ & $-0.206^{+0.067}_{-0.023}$ 
\\ \hline
BESIII~\cite{BESIIIX} & $ - $ & $0.706^{+0.088}_{-0.086}$ & 
$0.77^{+0.53}_{-0.49}$ & $-0.19^{+0.65}_{-0.63}$ 
\\ \hline
QCD sum rules~\cite{QCDSR1,QCDSR2} & $0.963\pm0.061$ & $0.708\pm0.047$ 
& $0.752\pm0.074$ & $ - $ 
\\ \hline
Cabibbo's model fit~\cite{Cabibbo} & $ - $ & $0.718\pm0.015$ 
& $1.066$ & $ - $ 
\\ \hline
Soliton model~\cite{soliton1,soliton2} & $ - $ & $0.68$ 
& $0.71$ & $ - $ 
\\ \hline
Quark model~\cite{quarkmodel1,quarkmodel2} & $ - $ & $0.724$ 
& $1$ & $ - $ 
\\ \hline
$1/N_{c}$ expansion~\cite{1/Nc} & $1.02\pm0.02$ & $0.729$ 
& $0.9$ & $ - $ 
\\ \hline
Lattice QCD~\cite{Lattic} & $0.9674\pm0.0047$ & $0.6902\pm0.0044$ 
& $-0.693\pm0.017$ & $-0.069\pm0.016 $ 
\\ \hline
Lattice QCD~\cite{LQCD} & $ - $ & $0.6902\pm0.0044$ 
& $ - $ & $ - $ 
\\ \hline
\end{tabular}
\end{table}

\sse{Decay branching ratios }

The amplitudes of $\Lambda \to p\,\ell^{-}\,\bar{\nu}_{\ell}~(\ell=e,\,\mu)$ contain some
independent mixtures of helicity components,  described by 
$h^{V(A)}_{\lambda, \lambda_{p}}$, where $\lambda$ and $\lambda_{p}$
represent the helicity components of the final baryon and $W$ propagator, respectively.
From Eq.~(\ref{Dwidth}), we can easily separate the integrals of the
longitudinal and transverse polarization asymmetries.
In Table \ref{Table3}, we show the decay branching ratios of 
$\Lambda \to p\,\ell^{-}\,\bar{\nu}_{\ell}~(\ell=e,\,\mu)$ 
in various LFQMs along with the BESIII data,  
where I and II represent our predictions without and with 
$f_{3}(g_{3})$ contributions, respectively. 
\begin{table}[htbp]
\caption{Branching fractions for the semileptonic decays 
$\Lambda\to p\ell^-\bar\nu_\ell$ ($\ell=e,\mu$).}
\vskip 0.2in
\label{Table3}
\begin{tabular}{|c||c|c|c|} \hline
 Result & ${\cal B}(\Lambda \to p\,e^{-}\,\bar{\nu}_{e})\times 10^{-4}$ & ${\cal B}(\Lambda \to p\,\mu^{-}\,\bar{\nu}_{\mu})\times 10^{-4}$ 
& $R^{e\mu}=\frac{{\cal B}(\Lambda \to p\,\mu^{-}\,\bar{\nu}_{\mu})}{{\cal B}(\Lambda \to p\,e^{-}\,\bar{\nu}_{e})}$
\\ \hline \hline
 I & $(8.32^{+2.63}_{-1.35}) $ & $(1.38^{+0.42}_{-0.21}) $ 
& $(0.166^{+0.073}_{-0.037}) $
\\ \hline
 II & $(8.32^{+2.63}_{-1.35}) $ & $(1.31^{+0.36}_{-0.20}) $ 
& $(0.158^{+0.066}_{-0.035}) $
\\ \hline
BESIII~\cite{BESIIIV,BESIIIX} & $8.16\pm0.26$ & 
$1.48\pm0.22$ & $0.181\pm0.028$ 
\\ \hline
LHCb~\cite{LHCb1} & $ - $ & 
$1.462\pm0.1$ & $0.175\pm0.012$ 
\\ \hline
PDG~\cite{PDG} & $8.34\pm0.14$ & 
$1.51\pm0.19$ & $0.181\pm0.022$ 
\\ \hline
QCD sum rules~\cite{QCDSR1,QCDSR2} & $7.72\pm0.64$ & $1.35\pm0.11$ 
& $0.174\pm0.016$ 
\\ \hline
QCD polynomial~\cite{QCDSR3} & $7.64^{+3.56}_{-0.94}$ & $1.5^{+0.80}_{-0.32}$ 
& $0.196^{+0.009}_{-0.012}$ 
\\ \hline
QCD z-expansion~\cite{QCDSR3} & $7.23^{+2.87}_{-0.94}$ & $1.26^{+0.45}_{-0.14}$ 
& $0.174^{+0.002}_{-0.005}$ 
\\ \hline
Cabibbo's model fit~\cite{Cabibbo} & $8.43\pm0.17$ 
& $ - $ & $ - $ 
\\ \hline
quark model~\cite{quarkmodel1,quarkmodel2} & $8.58$ 
& $1.49$ & $ 0.173 $ 
\\ \hline
Lattice QCD~\cite{Lattic} & $7.68\pm0.48$ & $1.33\pm0.16$ 
& $0.174\pm0.010$ 
\\ \hline
NLO-QCD~\cite{Nonst}& 
$ - $ & $ - $ & $0.153\pm 0.008$
\\ 
(Updated~\cite{LQCD})& 
$ - $ & $ - $ & $0.162\pm 0.004$
\\\hline
Lattice QCD~\cite{LQCD} & 
$ - $ & $ - $ & $0.163\pm 0.004$ 
\\ \hline
\end{tabular}
\end{table}
In particular, $f_{3}(g_{3})$ appears only in the time component 
of the helicity amplitude $H_{\frac{1}{2},t}^{V/A}$, 
specifically in the second term of Eq.~(\ref{Dwidth}). 
Therefore, $f_{3}(g_{3})$ gives a negligible contribution to the e-mode 
due to the tiny electron mass but does affect the $\mu$ mode.
It can be seen that the contribution with the inclusion of 
$f_{3}(g_{3})$ is about 6$\%$ less than that without it for the muon mode. 
In any case, our results are consistent with the BESIII data. 
A specific observable is the ratio of muonic to electronic decay rates, 
which is shown in Table~\ref{Table3}. 
As shown in the table, our results are lower than those of BESIII, 
primarily due to our lower muon decay rate.

Finally, we discuss the sensitivities of the input parameters to our results. 
In Table~\ref{Table3}, the central values of our results for the decay 
branching ratios correspond to the central ones of 
$\beta_{s[qq]}$, $\beta_{u[qq]}$, and $m_{[qq]}$ in Table~\ref{parameter}, 
while those of upper (lower) uncertainties are calculated by taking 
the upper (lower) values of $\beta$ and $m_{[qq]}$, respectively. 
Note that our results are based on fixed values of $m_q$ in Table~\ref{parameter}. 
We note that, different choices of these parameters will also affect the results. 
For instance, for $\Lambda \to p\,\ell^{-} \bar{\nu}_{\ell}$, 
if the diquark mass $m_{[qq]}$, which ranges from $0.4$ to $0.7$~GeV~\cite{diquark2}, 
changes without altering the baryon's $\beta$ value, 
a variation of diquark mass $m_{[qq]}$ by $\pm$0.1 GeV changes the 
branching ratios by approximately $\pm$11\%. 
Since this study assumes that the uncertainty of the $\beta$ factor is small, 
the uncertainty of the results is also small. 
Therefore, $m_{[qq]}$ is the main source of uncertainty.

\se{Conclusion}

We have calculated all six form factors of the $\Lambda \to p$ 
transition matrix element in the timelike region under the 
LF coordinate system and extended them to the entire physical region. 
These form factors are obtained based on the B-S formalism for the case $q^{+} > 0$ 
and include both valence and nonvalence contributions. In our calculations, 
by analyzing the form factors, 
as shown in Table II, the ratios $g_{1}/f_{1}$, $f_{2}/f_{1}$, and $g_{2}/f_{1}$ 
are given to be $0.708^{+0.012}_{-0.027},\, -0.872^{+0.014}_{-0.035}$, 
and $-0.206^{+0.089}_{-0.077}$, respectively. 
These form factors include nonvalence contributions and agree well 
with the existing data from the BESIII experiments 
as well as other theoretical estimates. 
In addition, for the results that include the contribution from $f_{3}(g_{3})$, 
we have found the branching ratios for the semileptonic decays of
$ \Lambda \to p\,e^{-}\,\bar{\nu}_{e}$ and $\Lambda \to p\,\mu^{-}\,\bar{\nu}_{\mu}$ 
being approximately 
$8.32\times 10^{-4}$ and  $1.31\times 10^{-4}$, which are close to the BESIII experimental results, respectively. 
However, we have obtained that 
$R^{e\mu}=\frac{{\cal B}(\Lambda \to p\,\mu^{-}\,\bar{\nu}_{\mu})}{{\cal B}(\Lambda \to p\,e^{-}\,\bar{\nu}_{e})}$=0.158, 
which is slightly smaller than the BESIII experimental value of $0.181$. 
The present analysis indicates that nonvalence contributions play an important role 
for a consistent light-front description of semileptonic baryon decays 
in the physical timelike region.
 Our framework may also be extended to 
other semileptonic baryonic transitions 
where nonvalence contributions become important in the physical timelike region.

\section*{Acknowledgments}

This work is supported in part by 
the National Natural Science Foundation of China (NSFC) under Grant No. 12547104.

\end{document}